# Functional Differences of MIP (Lens Fiber Major Intrinsic Protein) Between Animals and Birds

J. C. Phillips

Dept. of Physics and Astronomy

Rutgers University, Piscataway, N. J., 08854

Abstract

The major intrinsic protein (MIP) of the lens fiber cell membrane plays a role in lens biogenesis and maintenance.  Its polypeptide chains span the membrane six times, and the protein is naturally divided into two halves.  We use modern sequence analysis to identify differences between halves for humans (common to animals) and chickens (common to birds).

1. Introduction

An excellent survey of the evolution and functional differentiation of the two repeated halves of MIP proteins appeared decades ago [1].   It covered a very wide range of evolution, which was explained through statistical sequence analysis.  Here we focus on a much smaller topic, the titled differences.  We present a modern biophysical graphical sequence analysis that is similar to the one we recently used to discuss the functional differences between Piezo1 and Piezo 2 [2].

MIP is a 263 amino acid protein, whereas the Piezo proteins are ~ 10 times larger.  MIP has six transmembrane segments whereas the Piezo proteins have ~ 40 transmembrane segments.  However, MIP forms homotetramers [1], so the differences may not be so large.  Both families shape and stabilize membranes by mechanically acting on a large scale.

2. Methods

Our earlier analysis relied on Ψ(aa,W) hydropathic profiles, where Ψ(aa) measures the hydropathicity of each amino acid. Small changes in protein shapes are often driven by waves in water films. These water waves have been averaged linearly over sliding windows of width W.  (Data processing using sliding window algorithms is a general smoothing and sorting technique discussed online.)  A natural choice for W in transmembrane (TM) proteins is 21, as used by Uniprot in listing TM segments of Piezo1 and 2.  Wave motion has been an essential part of physics for centuries, but it is little used in molecular biology.  It was extremely useful in analyzing the evolution of Spike sequences and connecting them to CoV contagiousness [3-6].

3. Results

The hydropathic profiles with the KD scale (used in [2]) of human and chicken MIO are shown in Fig. 1.  The hydrophobic extrema (centers of transmembrane segments) are numbered 1-6.  The two halves (1-3) and (4-6) are similar, and are consistent with the description given in [1], of two nearly repeated halves.  The evolutionary differences between human and chicken are small.  Human 1-3 are more nearly level than chicken 1-3, whereas human 6 has become less hydrophobic than chicken 6.

A new feature of MIP sequences is apparent when we look at the hydrophilic extrema 7-12.  Hydroneutral is ~ 165, so the first half extrema 7-9 are still weakly hydrophobic ~ 170, whereas the second half extrema 10-12 at ~ 155 are clearly hydrophilic.  In other words, although the two halves appear to be nearly repeated inside the membrane, their dynamic behavior outside the membrane is quite different.  The second half is much more flexible than the first half.

Using "statistical analyses [1] suggested that the two halves of these proteins have evolved to serve distinct functions: the first half is more important for the generalized or common functions of these proteins, while the second half of these proteins is more differentiated to provide specific or dissimilar functions of the proteins". We reach a similar conclusion directly from Fig. 1, which replaces increased differentiation by increased flexibility.  In addition, these differences are enhanced near the C terminal edge 12.  Both N and C terminals face the cell cytoplasm [1].

A large difference occurs near the hydrophilic C terminal edge in the amino acid sequence of chicken 241-251, which is PPAAAPPPEPP, compared to human PDVSNGQPEVT. A peculiar aa repetition was identified in Piezo1 (dominated by E), while two such sequences were found in Piezo2, including the 21 aa sequence EKREEEEEKEEFEEERSREE. All three peculiar Piezo sequences are associated with deep hydrophilic edges. Here the dominant amino acid, occurring 7 times in chicken, but only twice in humans, is Proline. Proline is the only amino acid with a double connection to the peptide backbone, which is why Pro pairs have been so useful in stabilizing Spike vaccines [6-8]. Here they make the chicken N terminal more rigid. A BLAST search on similar sequences found many birds with 7 P, and a few lizards and turtles (only 5 P, instead of 7). Early in evolution the additional osmolarity stability was useful, especially for birds. The additional mechanical stability also supports and sustains rapid eye movement.

4. Discussion

Proteins are the most important and most studied examples of self-organized networks. The mechanical properties of such networks near critical points [9,10], have been extensively studied in network glass alloys, which (like proteins) are primarily covalently bonded [11]. The high accuracy of the biophysical results for functional evolution of CoV spikes has resulted from sequence analysis of the long, slender spike structures' immersion in water [3-6]. Similarly several of the functional differences between Piezo1 and 2 were explained by profiling sequential differences in their interactions with water [2]. In comparatively simple systems (like covalent glass alloys), it was possible to measure the elastic properties of the glass networks directly and plot them against force field constraints [11]. In this way critical compositions for elasticity and optimized glass formation were discovered.

In proteins this is seldom the case, even for proteins smaller than 300 amino acids. Here we have found striking differences between animal and bird MIP. Given their function as mechanotransducers, one expects that these two types could have significantly different elastic properties, and that both could be close to an evolutionary critical point. Similarly, criticality is a natural explanation for the hydrophilic differences between the two halves of MIP [1]. These differences suggest a model for MIP entry into membranes, as a snake with an N terminal head and a C terminal tail. Near criticality large changes in both elasticity and function occur [9-11].

The present graphical method of studying sequence differences is simple and direct. It might be another useful tool for in silico sequence assessment for antibody drug discovery [12].

Data availability: Amino acid sequences from Uniprot P30301 and P28238. Values of Ψ(aa) for the KD scale from Table 1 of [4].

Figure Caption

Fig 1. The hydropathic profiles of Human and Chicken MIP. The hydrophobic extrema 1-6 are centered on the transmembrane segments. The differences between the two hydrophobic halves 1-3 and 4-6 are small; they are much larger between the hydrophilic regions 7-9 and 10-12.

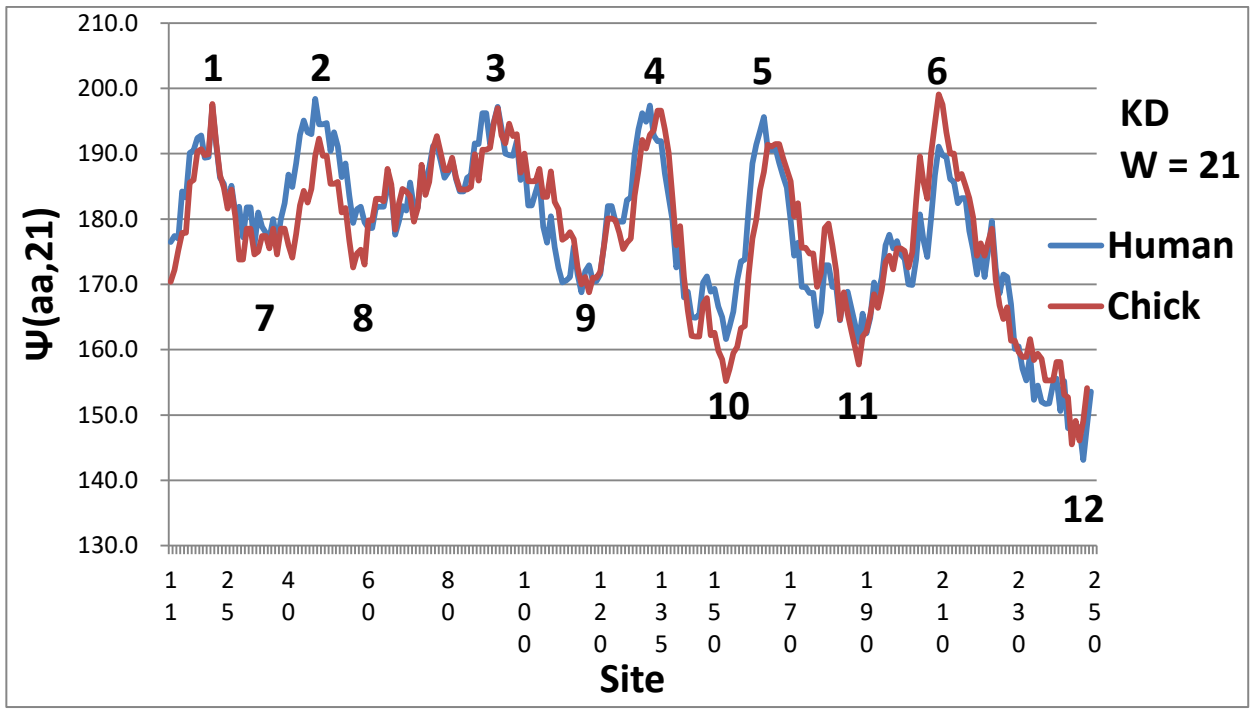

Fig. 1.